# Light-Induced Melting of Competing Stripe Orders without Introducing Superconductivity in $La_{2-x}Ba_xCuO_4$

S. J. Zhang[1,*], X. Y. Zhou[1], S. X. Xu[1], Q. Wu[1], L. Yue[1], Q. M. Liu[1], T. C. Hu[1], R. S. Li[1], J. Y. Yuan[1], C. C. Homes[2], G. D. Gu[3], T. Dong[1], and N. L. Wang[1,4,5,†]

[1]*International Center for Quantum Materials, School of Physics, Peking University, Beijing 100871, China*
[2]*National Synchrotron Light Source II, Brookhaven National Laboratory, Upton, New York 11973, USA*
[3]*Condensed Matter Physics and Materials Science Department, Brookhaven National Laboratory, Upton, New York 11973, USA*
[4]*Beijing Academy of Quantum Information Sciences, Beijing 100913, China*
[5]*Collaborative Innovation Center of Quantum Matter, Beijing, China*

The ultrafast manipulation of quantum material has led to many novel and significant discoveries. Among them, the light-induced transient superconductivity in cuprates achieved by melting competing stripe orders represents a highly appealing accomplishment. However, recent investigations have shown that the notion of photoinduced superconductivity remains a topic of controversy, and its elucidation solely through $c$-axis time-resolved terahertz spectroscopy remains an arduous task. Here, we measure the in-plane and out-of-plane transient terahertz responses simultaneously in the stripe-ordered nonsuperconducting $La_{2-x}Ba_xCuO_4$ after near-infrared excitations. We find that although a pump-induced reflectivity edge appears in the $c$-axis reflectance spectrum, the reflectivity along the $CuO_2$ planes decreases simultaneously, indicating an enhancement in the scattering rate of quasiparticles. This in-plane transient response is clearly distinct from the features associated with superconducting condensation. Therefore, we conclude the out-of-plane transient responses cannot be explained by an equivalent of Josephson tunneling. Notably, those pump-induced terahertz responses remain consistent even when we vary the near-infrared optical pump wavelengths and hole concentrations. Our results provide critical evidence that transient three-dimensional superconductivity cannot be induced by melting the competing stripe orders with pump pulses whose photon energy is much higher than the superconducting gap of cuprates.

## I. INTRODUCTION

The periodic modulations of charge density are widely observed in high-temperature superconducting cuprates (HTSCs) [1–3]. The intimate relationship between charge orders with superconductivity is considered key to understanding the mechanism of high-temperature superconductivity and has been the subject of intensive investigations in recent years [4–6]. In $La_{2-x}A_xCuO_4$ ($A$ = Ba, Sr), the charge order is nearly commensurate with the lattice and takes the form of one-dimensional stripes that coexist with spatially modulated antiferromagnetic orders [7], which is referred to as "stripe order." Notably, a 90° rotation of the stripe orientations in adjacent $CuO_2$ planes [illustrated in Fig. 1(a)] frustrates Josephson tunneling along the $c$ axis and hinders the formation of three-dimensional superconductivity, but is compatible with two-dimensional superconductivity within $CuO_2$ planes [8]. The in-plane two-dimensional superconductivity is also recognized as a pair density wave, where the order parameter of superconductivity undergoes periodic modulation in real space [9]. At the hole concentration of approximately $x = 0.125$, the stripe becomes "static" with a maximum coherence length, due to the pinning effect of the low-temperature tetragonal crystal structure [10]. That long-range stripe order leads to an anomalous suppression of the bulk, three-dimensional superconductivity [11], as shown in Fig. 1(b), while most HTSCs exhibit a dome-shaped doping-dependent superconducting phase boundary. The competition between the stripe order and three-dimensional superconductivity makes it possible to manipulate one order by tuning the other utilizing magnetic field [12], pressure [13–15], and even extremely low uniaxial stress [16,17].

[*]sjzh@pku.edu.cn
[†]nlwang@pku.edu.cn

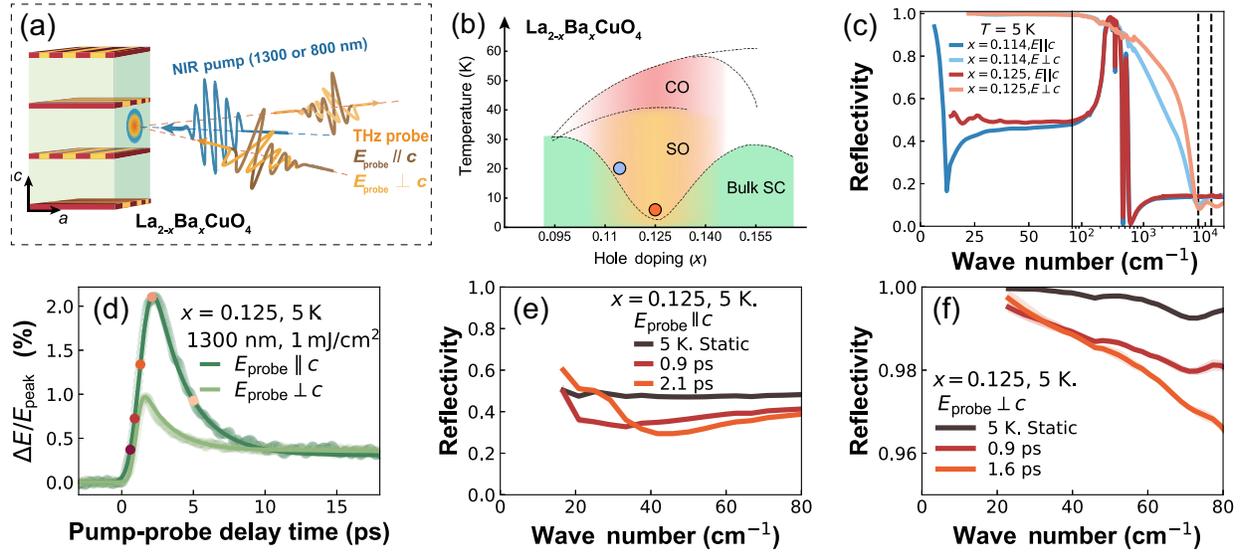

FIG. 1. (a) A schematic diagram of the optical-pump–terahertz-probe measurements on $La_{2-x}Ba_xCuO_4$ ($x = 0.114$ and $0.125$), in which the in-plane stripe order severely frustrates the interlayer Josephson coupling near the doping level $x = 0.125$. In the actual experiments, the terahertz probe beam is fixed to $s$ polarization, and the transient in-plane and out-of-plane terahertz responses are investigated by rotating the sample. (b) The phase diagram of $La_{2-x}Ba_xCuO_4$ (reproduced from Ref. [18]). At the doping level $x = 0.125$, superconductivity (SC) emerges at 2 K, charge stripe order (CO) at 54 K, and spin stripe order (SO) at 42 K, while for the $x = 0.114$ compound, $T_{SC} = 17$ K, $T_{CO} = 50$ K, and $T_{SO} = 40$ K. The colored circles indicate the transient terahertz responses reported here. (c) The broadband in-plane and out-of-plane reflectance spectra at 5 K. The black dashed lines indicate the pump wavelengths used for excitations: 1300 and 800 nm. (d) The decay of the relative change of the reflected terahertz electric field $\Delta E/E_{peak}$ in the $x = 0.125$ sample after excitations with 1300-nm pump polarized along the $c$ axis under an incident fluence of 1 mJ/cm$^2$ at 5 K. The colored dots indicate the delay times of the transient optical responses reported here. (e) A pump-induced edge appears in the $c$-axis reflectance spectrum of the $x = 0.125$ sample. "Static" indicates the delay time before the pump arriving. (f) The in-plane transient reflectivity decreases after excitations, which significantly differs from the effects of superconducting condensation.

Ultrafast light-matter interactions have recently become widely utilized for manipulating the macroscopic properties of quantum materials [19,20]. One promising method is to introduce or enhance an emergent phase by nonthermally suppressing its competing orders [21–28]. One of the most remarkable discoveries in this field is the observation of light-induced transient superconductivity in the stripe-ordered phase of HTSCs by melting the competing in-plane stripe orders [29–32]. To date, light-induced superconductivity in HTSCs was substantially evidenced by the appearance of a reflectivity edge in the $c$-axis terahertz reflectance spectrum accompanied by a divergent $1/\omega$ behavior near zero frequency in the imaginary part of optical conductivity $\sigma_2(\omega)$ after excitations. This has been interpreted as an analog to Josephson plasma reflectivity edge in the equilibrium superconducting state [33] introduced by suppressing the competing stripe order. However, a recent experimental result suggests that $c$-axis terahertz measurements cannot distinguish between the response of quasiparticles with extremely low scattering rates and that of Cooper pairs, which indicates that light-induced superconductivity is not an exclusive interpretation for those transient responses [34]. Furthermore, Katsumi et al. reported an absence of the nonlinear responses of $c$-axis Josephson tunneling in that light-induced transient state, suggesting a difference from the long-range-ordered equilibrium superconducting state [35].

Light-induced superconductivity in HTSCs remains a topic of controversy, as reviewed by Ref. [36]. Fortunately, the transient in-plane terahertz responses could help to explore the nonequilibrium state. In HTSCs, Cooper pairs are primarily confined within the $CuO_2$ planes, while the superfluid along the $c$ axis results from the Josephson tunneling of Cooper pairs [37]. It would be natural to conclude that once a transient superconductivity state is established along the $c$ axis, a simultaneous superconducting response should be observed within the $CuO_2$ planes as well. Otherwise, if the in-plane transient response is different from or even opposite to that of the superconducting state, the light-induced reflectivity edge appearing in the $c$-axis terahertz spectrum should not be attributed to light-induced transient superconductivity. Therefore, the controversy regarding light-induced superconductivity in HTSCs can potentially be resolved by monitoring the pump-induced changes of the in-plane electrodynamics.

Here, we investigate the transient terahertz responses parallel and perpendicular to the $c$ axis of $La_{2-x}Ba_xCuO_4$ (LBCO) induced by near-infrared optical pump. The stripe-order phases of LBCO single crystals at two different nominal hole concentrations $x = 0.114$ and $x = 0.125$ are examined. After the excitations by a near-infrared pump, although a reflectivity edge emerges along the $c$ axis, an obvious reduction of the in-plane reflectivity is observed simultaneously, which significantly differs from the effects of superconducting condensation. This decrease of the transient in-plane reflectivity signifies the excitation of quasiparticles and enhancement of the scattering rate occurs within the $CuO_2$ planes. Thus, we conclude the light-induced reflectivity edge in the $c$-axis terahertz spectrum cannot be attributed to Josephson tunneling of Cooper pairs, while the competing stripe order has been melted by ultrafast near-infrared excitations. Instead, it is likely due to an excitation of low-scattering quasiparticles.

## II. EXPERIMENTAL DETAILS

Figure 1(a) shows a schematic diagram of the optical-pump–terahertz-probe measurements, which are based on a Ti:sapphire regenerative amplifier laser system that produces 800-nm, 35-fs pulses at a 1-kHz repetition rate. The near-infrared optical pump beam is normally incident on the sample and set to either 800 or 1300 nm, the latter being achieved through an optical parametric amplifier pumped by the regenerative amplifier. The polarization of the pump beam is either parallel or perpendicular to the $c$ axis of LBCO. The terahertz probe beam is generated from 800-nm pulses via optical rectification, and its time-domain profile is detected by electro-optic sampling. The LBCO single crystals are grown using the traveling-solvent floating-zone method and cut at an angle of 90° relative to the $CuO_2$ planes to reveal a $c$-axis face, which is then finely polished to produce a mirrorlike surface. The transition temperatures for bulk superconductivity, charge stripe order, and spin stripe order are 2, 54, and 42 K for the $x = 0.125$ compound, and 17, 50, and 40 K for the $x = 0.114$ one [18], respectively, as shown in Fig. 1(b). The terahertz probe beam is set at a fixed $s$ polarization and an incident angle of 30° on the sample. By rotating the sample, we can measure the pump-induced transient terahertz responses parallel and perpendicular to the $c$ axis of LBCO.

Figure 1(c) presents the broadband reflectance spectra of both samples at 5 K. For the $x = 0.114$ compound, a sharp Josephson plasmon edge emerges near 10 cm$^{-1}$ in the $c$-axis reflectance spectrum, indicating the emergence of three-dimensional superconductivity. In contrast, the $x = 0.125$ compound exhibits incoherent $c$-axis charge dynamics with low reflectivity in the far-infrared region, while the in-plane dynamics resemble a metal with a plasmon edge near 1 eV. The photon energies of the near-infrared pump at 1300 and 800 nm are approximately 0.95 and 1.55 eV, respectively, as shown by the black dashed lines in Fig. 1(c).

## III. RESULTS OF $x = 0.125$ COMPOUND

We first present the transient responses of the $x = 0.125$ compound with a static stripe order after excitations with a 1300-nm pump polarized along the $c$ axis under an incident fluence of 1 mJ/cm$^2$ in the spin stripe-order phase at 5 K, where two-dimensional superconductivity has been established within the $CuO_2$ planes, while the out-of-plane coherence is impeded by in-plane static stripes [8,11]. Figure 1(d) shows the decay of the relative change of the reflected terahertz electric field $\Delta E/E_{\text{peak}}$. Upon excitation, the sample exhibits a maximum transient response within 1.6 ps for the $CuO_2$ planes and 2.1 ps for the $c$ axis. The decay procedure can be well fitted by a single-exponential decay function, with a decay time of approximately 2 ps, as shown by the solid lines in Fig. 1(d). This decay time is much faster than that in the superconducting state of HTSCs [38–41] but consistent with previous results in the stripe-ordered phase [30–32].

The 1300-nm pump polarized along the $c$ axis has a penetration depth of approximately 1 μm, while the terahertz probe polarized along the $CuO_2$ planes penetrates approximately 0.3 μm and over 10 μm along the $c$ axis. Therefore, the measured out-of-plane $\Delta E/E_{\text{peak}}$ reflects a combination of the pump-induced transformed surface layer and a large proportion of unexcited components. This requires a multilayer model to extract the transient optical constants within the transformed regime [34]. Figure 1(e) illustrates the calculated reflectivity along the $c$ axis at several pump-probe delay times. In comparison to the insulatinglike response of the equilibrium state, which is a constant of approximately 0.5, a pump-induced reflectivity edge appears in the terahertz regime. This edge bears a resemblance to the Josephson plasmon edge depicted in Fig. 1(c), albeit with a more gentle slope. Meanwhile, the in-plane transient reflectivity decreases significantly after excitations as shown in Fig. 1(f), which significantly differs from the increase of reflectivity approaching the equilibrium superconducting state [42].

### A. Out-of-plane transient responses

To clarify the origin of the light-induced reflectivity edge appearing in the $c$-axis terahertz spectroscopy, we investigate the out-of-plane optical conductivity spectrum at 5 K. In the equilibrium state, the $c$-axis transport exhibits nearly insulating behavior with minimal contribution from carriers. Figure 2(a) displays the energy-loss function Im$[-1/\tilde{\varepsilon}(\omega)]$. In the equilibrium state, the $c$-axis energy-loss function is a constant and close to zero. After excitations, a wide peak in the loss function appears at a very low frequency and progressively moves toward higher frequencies during 0.6 to 2.1 ps, thereby signifying the emergence of a reflectivity edge in Fig. 1(e). Figure 2(b) shows the corresponding real part of the conductivity $\sigma_1(\omega)$. After 0.9 ps, the pump-induced change of $\sigma_1(\omega)$ spectral weight primarily happens near zero frequency and gradually evolves as a Drude-like peak, unambiguously indicating the emergence of quasiparticles along the originally insulatinglike $c$ axis. At 2.1 ps, the reflectivity edge attains its most significant manifestation. The distinct Drude-like peak in $\sigma_1(\omega)$ and the kink near 25 cm$^{-1}$ in the imaginary part of the conductivity $\sigma_2(\omega)$ in Fig. 2(c) indicate that the reflectivity edge is a consequence of pump-induced quasiparticles, rather than Cooper pairs. The scattering rate of those quasiparticles can be deduced by the full width at half maximum of the Drude peak in $\sigma_1(\omega)$ and the low-frequency turning point of $\sigma_2(\omega)$. Subsequently, the sample quickly decays to the insulatinglike response of the equilibrium state.

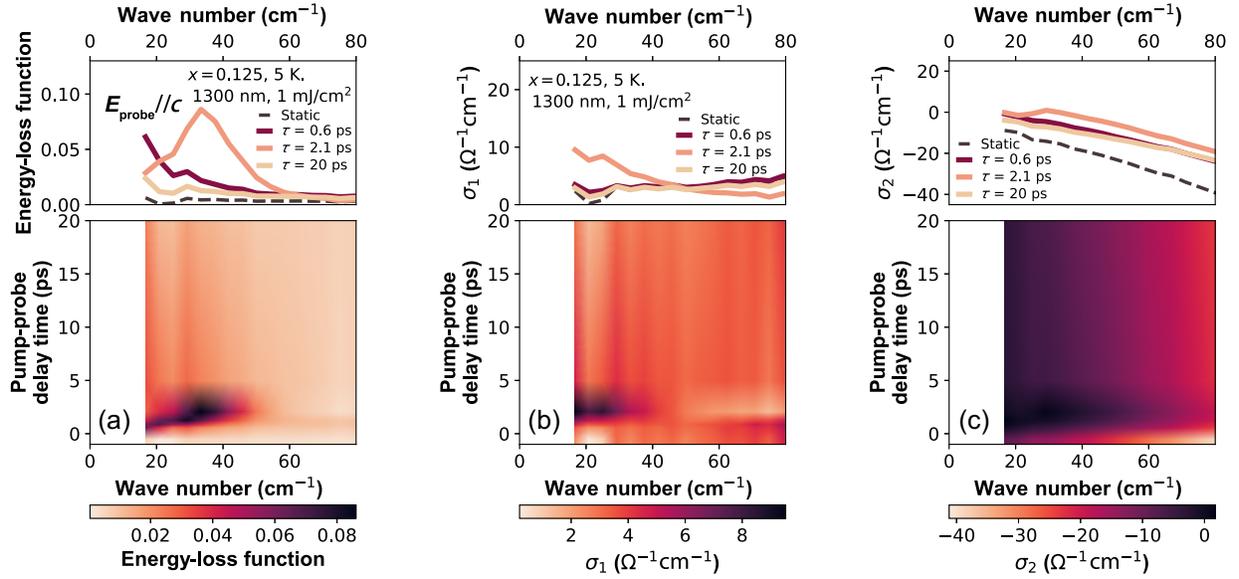

FIG. 2. Transient terahertz responses along the $c$ axis of the $x = 0.125$ sample at 5 K after excitations with 1300-nm pump polarized along the $c$ axis under an incident fluence of 1 mJ/cm$^2$. (a) displays a pump-induced emergence of a peak in the energy-loss function, indicating a transient reflectivity edge after excitations. The (b) real and (c) imaginary parts of the transient optical conductivity. At 0.6 ps, $\sigma_2(\omega)$ near zero frequency increases accompanied by $\sigma_1(\omega)$ remaining unchanged. The low-frequency measurement range poses a challenge in determining whether the transient response is contributed by quasiparticles with an extremely low scattering rate or superfluid carriers. After 0.9 ps, the pump-induced quasiparticles are evidenced by a Drude-like peak in $\sigma_1(\omega)$.

However, the interpretation of the response at 0.6 ps, which is associated with the rising process in Fig. 1(c) and marked by a reflectivity edge and an increase in $\sigma_2(\omega)$ near zero frequency while $\sigma_1(\omega)$ remains constant, is still being debated. That phenomenon has not been previously observed in La$_{1.875}$Ba$_{0.125}$CuO$_4$ [30]. However, it has been reported in La$_{1.675}$Eu$_{0.2}$Sr$_{0.125}$CuO$_4$ with a static stripe order [29,43], La$_{1.885}$Ba$_{0.115}$CuO$_4$ [30,31,44], and underdoped YBa$_2$Cu$_3$O$_{6-x}$ [45], albeit with varying upward slopes in the reflectivity edge. Previously, that phenomenon was attributed to light-induced superconductivity and was fitted with a two-fluid model that included a superconducting component. Nevertheless, similar $c$-axis far-infrared spectra exhibiting metallic behavior have been observed in the equilibrium normal state of overdoped La$_{1.82}$Sr$_{0.18}$CuO$_4$ [46], La$_{1.73}$Nd$_{0.12}$Sr$_{0.15}$CuO$_4$ [47], and underdoped YBa$_2$Cu$_3$O$_{6.7}$ as well [48], which results from carriers with an extremely low scattering rate. Because of the measurement range limitations, the true origin of the transient reflectivity edge cannot be determined by the $c$-axis time-domain terahertz spectroscopy.

### B. In-plane transient responses

To investigate the possibility of light-induced three-dimensional superconductivity, it is necessary to further study the in-plane transient responses after excitations. Figure 3(a) presents the equilibrium temperature-dependent in-plane optical properties in the terahertz range of the $x = 0.125$ compound. As the temperature decreases, the in-plane reflectivity increases monotonically. Below 16 K, two-dimensional superconductivity has already been established, yet zero-resistance transport along the $c$ axis is still hindered [11]. As a result of superconducting condensation, the in-plane reflectivity at 5 K approaches nearly unity, as depicted in the upper panel of Fig. 3(a), while the out-of-plane reflectivity remains consistent with the normal state, as illustrated in Fig. 1(c). Regarding the in-plane optical conductivity, $\sigma_1(\omega)$ exhibits an initial increase in the terahertz regime as warming up from 5 K, as depicted in the middle panel of Fig. 3(a). This increase is attributed to the emergence of a Drude component due to an increased quantity of normal carriers. As the temperature rises, $\sigma_1(\omega)$ follows a Drude-like response with a constant plasma frequency and an increasing scattering rate evidenced by a progressively broadening peak centered at zero frequency in $\sigma_1(\omega)$ [42]. However, within the limited terahertz-frequency regime, $\sigma_1(\omega)$ remains nearly flat.

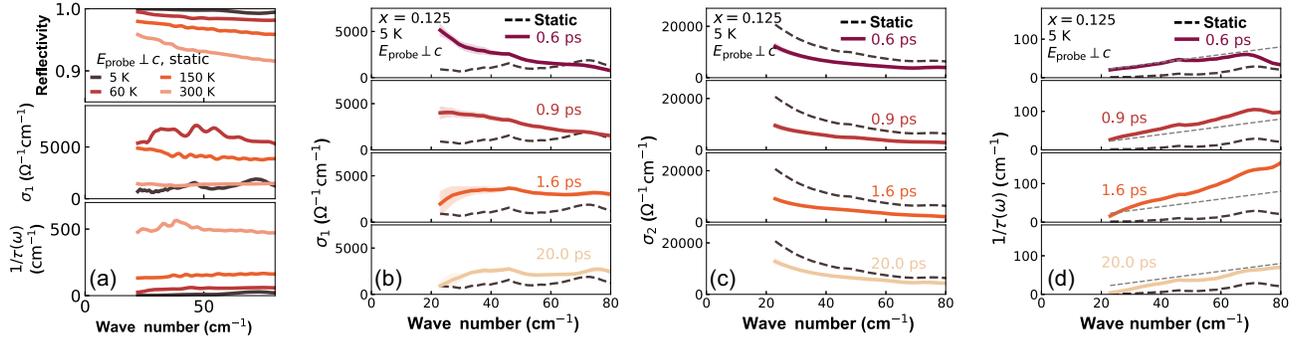

FIG. 3. Terahertz responses along the $CuO_2$ planes of the $x = 0.125$ sample. (a) Temperature-dependent terahertz responses in the equilibrium state. The (b) real and (c) imaginary parts of the transient optical conductivity after excitations with 1300-nm pump polarized along the $c$ axis under an incident fluence of 1 mJ/cm$^2$ at 5 K. (d) The frequency-dependent scattering rate in the extended Drude model, with the gray dashed line plotting $1/\tau(\omega) = \omega$, above which the transport is highly dissipative and far away from a Landau-Fermi liquid. The increase of $\sigma_1(\omega)$ and $1/\tau(\omega)$ accompanied by a decrease in $\sigma_2(\omega)$ indicate an excitation of quasiparticles and enhancement of transport dissipation. There is no evidence for the emergence of transient superconducting condensate.

The analysis of in-plane electrodynamics in LBCO with strongly correlated electron systems can be carried out using an extended Drude model [49]. The frequency-dependent scattering rate of quasiparticles is

$$\frac{1}{\tau(\omega)} = \frac{\omega_p^2}{4\pi} \text{Re}\left[\frac{1}{\tilde{\sigma}(\omega)}\right],$$

where $\omega_p$ is the in-plane plasma frequency and $\tilde{\sigma}(\omega)$ the complex optical conductivity. As warming occurs, the in-plane scattering rate exhibits a monotonic rise in the terahertz regime, as shown in the lower panel of Fig. 3(a). Consequently, the pump-induced change of in-plane reflectivity and scattering rate can serve as the optical indicators of two-dimensional superconductivity. We also emphasize that the terahertz probe polarized along the $CuO_2$ planes penetrates less deeply (approximately 0.3 μm) than the near-infrared optical pump (approximately 1 μm). Consequently, it is possible to extract the in-plane transient responses without calculating with the multilayer model.

The transient in-plane reflectivity of the $x = 0.125$ compound after excitations has been already displayed in Fig. 1(f). A distinct decrease in reflectivity is clearly observed, significantly distinguished from the behavior associated with superconducting condensation. Figure 3(b) shows the corresponding transient optical conductivity. At 0.6 and 0.9 ps, a light-induced Drude-like peak appears in $\sigma_1(\omega)$. This suggests that the pump pulses cause a rapid increase in the scattering rate of the quasiparticles within the $CuO_2$ planes, and may also be related to the partial destruction of the in-plane two-dimensional superconducting condensate [50]. At 1.6 ps, the in-plane transient change reaches its maximum and $\sigma_1(\omega)$ demonstrates an overall increase, rather than a distinct zero-frequency Drude-like peak. In the $x = 0.125$ sample, a charge or spin order induces a density-wave gap causing a transfer of spectral weight from low frequency to high frequency in the equilibrium state [42]. The significantly increased transient spectral weight in $\sigma_1(\omega)$ at 1.6 ps is also partly due to the destruction of the density-wave gap, leading to the transfer of spectral weight back from higher frequencies to the terahertz regime. This is additionally supported by the fact that the in-plane stripe order is shown to instantaneously melt by ultrafast laser in time-resolved x-ray diffraction and scattering measurements [51,52]. The transient responses gradually decay with pump-probe delay time and stabilize over tens of picoseconds, which may also be related to a light-induced transformation of the low-temperature tetragonal crystal structure [53]. The in-plane scattering rate is significantly enhanced after excitations, even assuming $\omega_p$ remaining consistent with that in the equilibrium state, as shown in Fig. 3(c), distinguished from the transition introduced by superconducting condensation as well.

## IV. RESULTS OF $x = 0.114$ COMPOUND

To enable a more direct comparison with the previously reported light-induced superconductivity in $La_{1.885}Ba_{0.115}CuO_4$ [30], we replicate the measurements on the $x = 0.114$ compound. After excitations of 800 nm at an incident fluence of 2 mJ/cm$^2$, the reflected $\Delta E/E_{peak}$ exhibits a rapid decay within 2 ps, as illustrated in Fig. 4(a). In the $c$-axis reflectance spectrum, a pump-induced reflectivity edge emerges and grows more pronounced with higher pump fluence, as shown in Fig. 4(b). Simultaneously, the in-plane reflectivity experiences a clear decrease after the excitations, as depicted in Fig. 4(c). The transient responses of the $x = 0.114$ compound, including the pump-induced decay processes and changes in the reflectance spectra, bear a strong resemblance to those observed in the $x = 0.125$ one, as reported earlier. Crucially, the transient reduction of in-plane reflectivity in the $x = 0.114$ compound also distinguishes itself from the behavior associated with superconducting condensation.

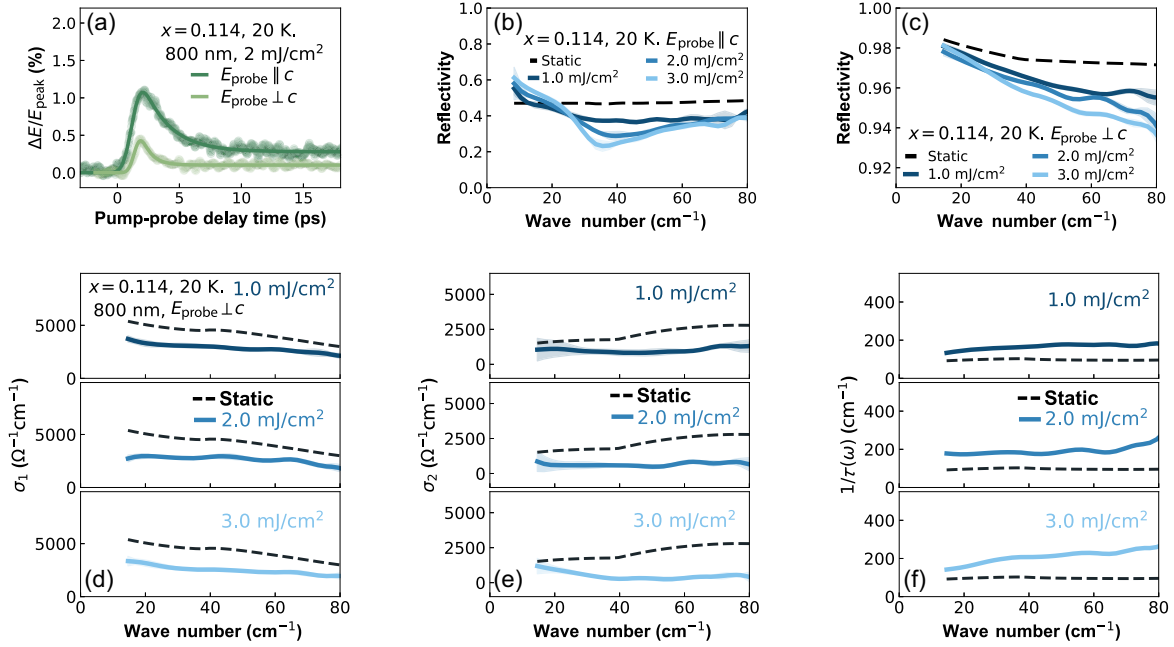

FIG. 4. Transient terahertz responses of the $x = 0.114$ sample after excitations with a 800-nm pump polarized along the $c$ axis at 20 K. (a) The decay of $\Delta E/E_{\mathrm{peak}}$ after excitations by an incident fluence of 2 mJ/cm$^2$, which is qualitatively similar to that observed in the $x = 0.125$ sample. (b) A pump-induced edge appears in the $c$-axis reflectance spectrum, becoming more significant with increasing pump fluence. (c) The in-plane transient reflectivity decreases after excitation. The (d) real and (e) imaginary parts of the transient in-plane optical conductivity. (f) The frequency-dependent scattering rate. All transient properties are taken at the time delay with a maximum response.

We then investigate the in-plane optical conductivity spectra of the $x = 0.114$ compound. In the equilibrium state at 20 K, $\sigma_1(\omega)$ has an obvious Drude component and exhibits a high value in the terahertz regime, reaching close to 5000 $\Omega^{-1}$ cm$^{-1}$ near 20 cm$^{-1}$, shown as the black dashed curves in Fig. 4(d). Meanwhile, $\sigma_2(\omega)$ shows limited values as presented in Fig. 4(e). This is significantly different from the two-dimensional superconducting behavior in the $x = 0.125$ compound at 5 K in Fig. 3, where $\sigma_1(\omega)$ is approximately 1000 $\Omega^{-1}$ cm$^{-1}$ near 20 cm$^{-1}$ and $\sigma_2(\omega)$ exhibits a nearly $1/\omega$-dependent divergence. The absence of superconducting behavior in the $x = 0.114$ sample at 20 K suggests a probable fluctuating in-plane superconductivity. After excitation, both in-plane conductivity components $\sigma_1(\omega)$ and $\sigma_2(\omega)$ exhibit synchronous reduction in the $x = 0.114$ sample, accompanied by a boosted $1/\tau(\omega)$, as seen in Figs. 4(d)–4(f). Those pump-induced changes remain qualitatively identical but become more pronounced with an increase in pump fluence from 1 to 3 mJ/cm$^2$. The transient changes of in-plane reflectivity $\sigma_2(\omega)$ and $1/\tau(\omega)$ collectively signify a pump-induced enhancement in the scattering rate of quasiparticles within the CuO$_2$ planes. Consequently, the reduction of in-plane $\sigma_1(\omega)$ is attributed to a further broadening of the Drude component. Therefore, we conclude that the pump-induced reflectivity edge along the $c$ axis of the $x = 0.114$ sample does not arise from Josephson current either, due to the absence of pump-induced in-plane superconductivity.

To assess pump wavelength-dependent in-plane transient responses, we examine the pump-induced changes in the $x = 0.114$ compound at 20 K after excitations of 800- and 1300-nm pumps. The corresponding penetration depths are approximately 0.4 and 1 μm, respectively. Figures 5(a) and 5(b) illustrate transient reflectivity and frequency-dependent scattering rate $1/\tau(\omega)$. After exposure to different pump wavelengths, in-plane transient responses consistently demonstrate reduced reflectivity and increased dissipation, indicating no substantial distinctions between the transient states induced by different pump wavelengths. Notably, a slightly more pronounced change is observed with the 800-nm pump compared to the 1300-nm one, both applied at an incident pump fluence of 2 $mJ/cm^2$. This discrepancy may stem from a higher quantity of absorbed photons per unit volume for the 800-nm pump [54]. In scenarios where the density of absorbed photons is almost identical, such as the 800-nm pump at 1 $mJ/cm^2$ and the 1300-nm pump at 2 $mJ/cm^2$, the transient responses are quantitatively identical. Moreover, these transient responses are quantitatively insensitive to variations in the polarization of the pump beam, as illustrated in Figs. 5(c) and 5(d). In conclusion, the pump-induced transient responses appear indifferent to specific near-infrared optical pump beams.

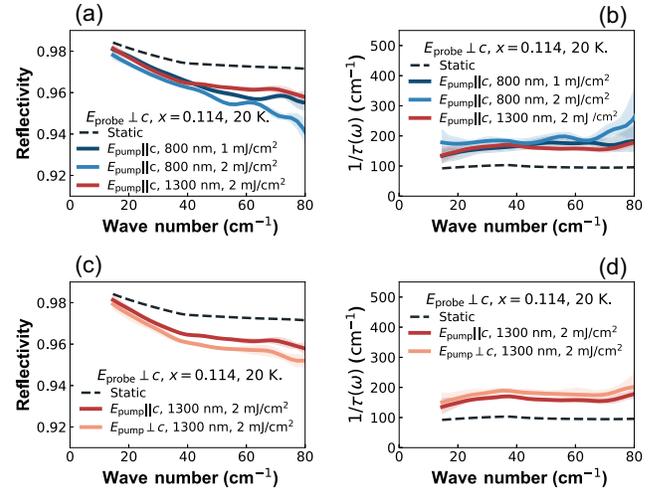

FIG. 5. The in-plane transient terahertz responses of the $x = 0.114$ sample are investigated under various optical excitation conditions. (a) and (b) present the results dependent on the pump wavelengths. (c) and (d) demonstrate that altering the pump polarization has no significant impacts.

## V. DISCUSSION

Based on those in-plane transient optical responses in both the $x = 0.114$ and $x = 0.125$ compounds after excitations, we conclude that although there is spectral evidence of stripe-order destruction, no transient in-plane superconducting condensate is observed at any delay times. In contrast, the near-infrared pump, with photon energy significantly much higher than both the superconducting and density-wave gaps in LBCO, excites quasiparticles in the $CuO_2$ planes, increasing transport dissipation and suppressing in-plane superconducting condensate by injecting an overwhelming amount of energy into the system. Those in-plane transient responses prevent us from attributing the pump-induced transient reflectivity edge observed along the $c$ axis depicted in Figs. 1(e) and 4(b) to the emergence of three-dimensional superconductivity. This is because in-plane superconducting condensation is a prerequisite for Josephson coupling between adjacent $CuO_2$ planes. Therefore, we conclude that the $c$-axis reflectivity edge arises from an excitation of low-scattered quasiparticles.

Then, the key issue is to understand how carriers can conduct with such a small scattering rate after excitation. On the one hand, previous investigations have reported a rather complicated charge conduction along the $c$ axis, whose hopping integral is strongly in-plane momentum dependent, being zero at nodal direction and maximum at the antinodal direction for the simple tetragonal cuprate system. Furthermore, the scattering rates of carriers contributing to the $c$-axis conductivity could be different at different regions of the Fermi surface or Fermi arc [55–57]. On the other hand, earlier beliefs regarding a very large quasiparticle scattering rate along the $c$ axis have been challenged by the subsequent experimental observations of clear metalliclike behaviors in the equilibrium normal states of overdoped $La_{1.82}Sr_{0.18}CuO_4$ [46], $La_{1.73}Nd_{0.12}Sr_{0.15}CuO_4$ [47], and underdoped $YBa_2Cu_3O_{6.7}$ [48]. Therefore, we conclude the observed light-induced reflectivity edge along the $c$ axis results from a photodoping effect by introducing additional dopants into the $CuO_2$ planes. This transfers the sample into the overdoped regime with ultrafast near-infrared pump pulses, supported both theoretically and experimentally in HTSCs [58,59].

## VI. CONCLUSION

In conclusion, we report the in-plane and out-of-plane transient terahertz responses of the stripe-ordered $La_{2-x}Ba_xCuO_4$, induced by near-infrared optical pump pulses. Even though the near-infrared excitations can induce a reflectivity edge along the $c$ axis, reminiscent of a Josephson plasmon edge in the equilibrium superconducting state, the in-plane reflectivity undergoes a simultaneous decrease, definitely opposing an in-plane superconducting condensation. This implies that the pump-induced reflectivity edge along the $c$ axis cannot be attributed to the emergence of three-dimensional superconductivity but indicates the emergence of quasiparticles with very low scattering rate, possibly due to the photodoping effect by transiently melting the in-plane stripe orders. Our results also show that these pump-induced transient responses are independent of both the near-infrared pump wavelengths and hole concentrations. Consequently, transient three-dimensional superconductivity cannot be induced by melting competing stripe orders using pump pulses with a photon energy much higher than the superconducting gap of HTSCs.

## ACKNOWLEDGMENTS

We gratefully acknowledge J. M. Tranquada for illuminating discussions. This work was supported by National Natural Science Foundation of China (Grant No. 11888101) and the National Key Research and Development Program of China (Grant No. 2022YFA1403901). Work at Brookhaven National Laboratory was supported by the Office of Science, U.S. Department of Energy under Contract No. DE-SC0012704. S. J. Z. was also supported by Natural Science Foundation of China (Grant No. 12304184).

## APPENDIX: IMPACT OF ESTIMATED PENETRATION DEPTH OF PUMP BEAM ON THE CALCULATED OUT-OF-PLANE RESPONSES

For HTSCs, before entering the three-dimensional superconducting state, the penetration depth of a terahertz probe beam along the $c$ axis of crystals significantly exceeds that of the near-infrared optical pump one. Specifically, the former extends beyond 10 μm, while the latter typically remains within 1 μm. This implies that the measured raw data, i.e., out-of-plane relative change of the reflected terahertz electric field $\Delta E/E_{\mathrm{peak}}$, reflects a combination of the pump-induced transformed surface layer and a substantial proportion of unexcited components. Consequently, a multilayer model is essential to extract the transient optical constants within the pumped regime [34].

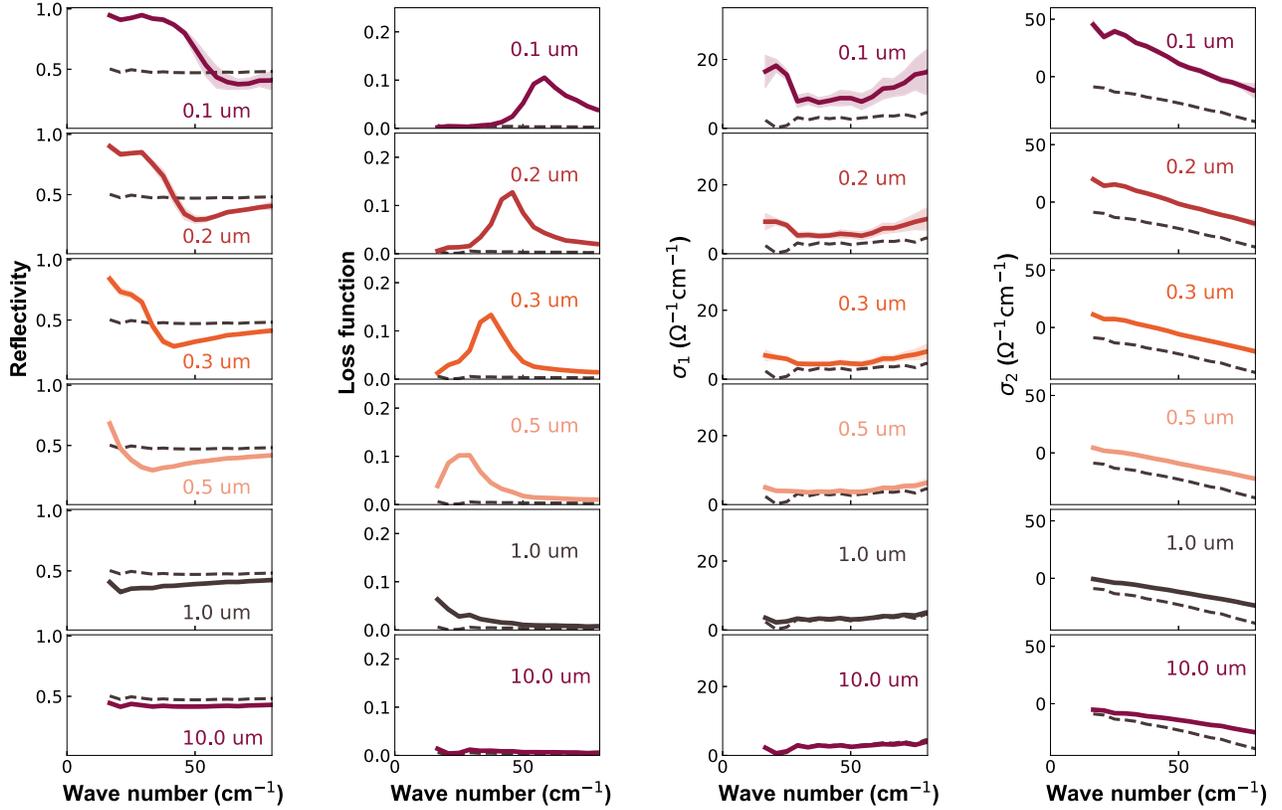

FIG. 6. The out-of-plane transient responses of the $x = 0.125$ sample calculated with the multilayer model. This calculation utilizes identical raw data acquired at 0.6 ps after excitations with a 1300-nm pump polarized along the $c$ axis under an incident fluence of 1 mJ/cm$^2$ at 5 K, but by substituting different penetration depths of the pump pulse into the calculation. The calculated transient responses are highly sensitive to the chosen penetration depth when it is shorter than 1 μm.

In the multilayer model, the penetration depth of the pump beam, derived from the corresponding complex refractive index, stands as a crucial parameter. Here, we demonstrate this by computing the transient $c$-axis responses of the $x = 0.125$ sample. We utilize identical raw data but substitute different penetration depths of the pump beam into the multilayer model. The calculated results are depicted in Fig. 6, which underscore that the calculated transient out-of-plane responses exhibit high sensitivity to the value of penetration depth, particularly when it is less than 1 μm. A shorter penetration depth induces a sharper and blueshifted transient reflectivity edge, with $\sigma_2(\omega)$ approaching a divergent behavior near zero frequency. Furthermore, it is crucial to recognize that there are two distinct definitions of penetration depth, namely, $d(\omega) = (c/\omega \, \mathrm{Im}[\tilde{N}(\omega)])$ and $d(\omega) = (c/2\omega \, \mathrm{Im}[\tilde{N}(\omega)])$. The choice between those two definitions depends on whether the depth is over which the electric field or the intensity of optical beam decays in the material by a factor of $1/e$. In this article, we opt for the former, namely, the depth associated with the decay of the electric field. In summary, both the definition and quantity chosen can significantly impact the calculated transient responses along the $c$ axis.

The out-of-plane transient terahertz responses of stripe-ordered La$_{2-x}$A$_x$CuO$_4$ ($A =$ Ba, Sr) after the excitation of a near-infrared optical pump have been previously reported by three research groups [30–32]. In line with our observations, all three groups documented a pump-induced reflectivity edge along the $c$ axis, albeit with significant variations in sharpness. This divergence could potentially stem from differences in the estimated value of the penetration depth of the pump beam.

The in-plane metallic responses enable the terahertz probe to penetrate within approximately 0.5 μm, a depth comparable to the penetration depth of optical pump pulses. This alignment eliminates the penetration mismatch issue, guaranteeing that in this scenario, the in-plane transient optical properties can be derived from the raw measured data without resorting to a multilayer model. Consequently, delving into the in-plane transient responses of $La_{2-x}A_xCuO_4$ has the potential to yield more precise insights into the light-induced state.